# ANTI-LATINX COMPUTATIONAL PROPAGANDA IN THE UNITED STATES

Claudia Flores-Saviaga, Saiph Savage.

HCI Lab, West Virginia University
Universidad Nacional Autonoma de Mexico
(UNAM)

# Executive Summary

How did social media discussions around the Latino community evolve during the 2018 US midterm elections? How might this affect the involvement of Latinos in and around the election? As the Latino community is the second largest ethnic group in the US, understanding how Latinos are discussed, mentioned (targeted) on social media during US elections becomes crucial. This working paper starts to answer these questions through a data analysis on one the most prominent and popular social media platforms for political discussion: Reddit. We collected Reddit posts mentioning Latinos and the US midterm elections from September 24th, 2017 to September 24th, 2018. We analyzed people's posting patterns over time and the digital traces of the individuals pushing the majority and most popular content. Our research hints that there are three main type of actors discussing Latinos and politics on Reddit: (1) individuals posting about how "president Trump is anti-Latinos"; (2) people who simply posted news stories about Latinos and politics; and (3) political trolls focused on mobilizing Latinos to vote Republican (pro-Trump). The political trolls shared stories of how Latinos across the US were also supporting Trump and how Trump's policies against illegal immigrants from Latin America were beneficial to the US, especially Latinos who were US Citizens.

Our research highlights current data voids that exists in online discussions surrounding Latinos and the US elections. We observed a lack of neutral actors engaging Latinos in political topics. It appeared that it was more the extremist voices, i.e., the individuals operating within subreddits who identify themselves as political trolls, who were creating the most political content about Latinos and a lack of neutral actors participating in the discussions. We finish by discussing the possible dangers of these data voids, especially their ties to misinformation and recommendations to increase the involvement of the Latino community in future US elections.

# Contents

# 1. Introduction

In the 2008 and 2012 US presidential campaigns, there was a belief that digital tools could enhance democracy by expanding citizen empowerment and engagement (J. Tucker et al., 2018). However, after the 2016 US presidential campaign new concerns arose highlighting that social media may have been weaponized to undermine democracy (Persily, 2017). Current research has been investigating how social media was used as a tool to share computational propaganda for spreading disinformation or causing social disruptions (Woolley & Howard, 2017; Wooley & Howard, 2016).

While most current research has focused on how disinformation is targeting populations, little is known about how the online discussions about the Latino community is evolving on social media. Understanding these online discussions is important to ensure that the Latino community is getting involved in political discussions prior the elections and therefore, ensure fair elections. According to Pew Research Center, 27.3 million Latinos were eligible to vote in 2016, a share greater than any other ethnic group of voters, representing 12% of all eligible voters (Krogstad, 2016).

Online social media networks have enabled people to read and share news, discuss important events and engage in political discussions (Yaqub, Chun, Atluri, & Vaidya, 2017). Recent research concluded that Reddit was among the platforms that played a very important role in the dissemination of information during the US presidential elections of 2016 (Kreiss & McGregor, 2018; Roozenbeek & Palau, 2017). For this reason, we focus on Reddit to study how Latinos are targeted (mentioned) in online political discussions surrounding the 2018 US midterm elections using one year of posts, from September 24, 2017 to September 24, 2018.

We characterize the content that people post about Latinos during the 2018 midterm elections using analytic techniques similar to those used for examining discourse between politicians and their audiences (Flores-Saviaga, 2018; Howard, Savage, Saviaga, Toxtli, & Monroy-Hernández, 2016; Larsson & Moe, 2012). Our aim is to explore the context under which Latinos and electoral content gained attention over time and to investigate the behavioral patterns of the people pushing such content.

Through our analysis, we uncover that the conversations on the different subreddits came mainly from extremist voices, such as pro-Trump political trolls. We call political trolls to those participating in the the alt-right movement that burst onto the national political scene in 2015 (Bokhari A., 2016). It includes white nationalists and disillusioned right-wing dissidents who capitalized on the anti-immigration and anti-establishment campaign themes of Donald Trump to thrust its ideas into the political mainstream during the 2016 US elections. They utilized memes–and like Trump, the movement attracted attention and visibility through provocations and sensationalism (Heikkilä, 2017). Previous research has discover them on social media platforms such as Reddit (Merrin, 2019; Flores-

Saviaga, 2018; Jamieson, 2018).

In this research we uncover that political trolls had numerous strategies for engaging people in topics around Latinos and the elections, such as creating posts where people could directly talk with politicians (AMAs) and having mega-threads where people could have deep discussions around the topics they cared about. Our investigation suggests that extremist voices are the main ones covering political issues regarding Latinos on reddit. Our data analysis starts to expose the extent to which this is happening. While news media had covered the lack of interest of different actors to cover political events involving the Latino community (Bosquez, 2018), the data analysis in this working paper reveals that in effect there are extremist voices discussing Latino topics that appear to be louder than their supporters and they seem to be filling current data void that exists in online discussions surrounding Latinos and the US elections.

The literature has defined data voids to be "search terms for which the available relevant data is limited or non-existent" (). Usually, data voids emerge because they are related to concepts that people rarely search for or that people generally do not generate content for. Data voids can be problematic because they can be potentially exploited by those with ideological, economic, or political agendas. In this case we are identifying that political trolls appear to be occupying a data void that exists about Latinos in the midterm elections. Political trolls appear to be using this data void to push their own content and narrative around Latinos and the midterm elections. All of this likely to have more people exposed to their ideas. For instance, if someone interested in the topic searched for information on reddit, they would be more likely to stumble on the content generated by political trolls along with the surrounding stories they are pushing, than content from more neutral sources (as it is almost nonexistent.)

We finish by discussing the combined implications of our findings, with particular concern for the lack of neutral actors engaging Latinos in political topics.

## 2. Literature Review

**Disinformation and its Effects in Democracy**

Social media allows anyone to initiate public political debates easily, but it can also give a voice to extremist voices and actors seeking to manipulate the political agenda in their own financial or political interest by spreading disinformation (J. Tucker et al., 2018; J. A. Tucker, Theocharis, Roberts, & Barberá, 2017). The spread of political disinformation and propaganda online is considered to have negative societal outcomes (Marwick & Lewis, 2017). Previous research has shown that misinformation can be amplified in communities where people with a similar point of view coincide and can be accepted as true partly due to a lack of dissenting voices that could challenge it, creating a *filter bubble* effect (Pariser, 2011). This brings as a consequence a society that is polarized (Sunstein, 2018). Minorities are most often one of the groups most susceptible to disinformation (J. Tucker et al., 2018). Disinformation can lead to distortions in the collective public opinion about minority groups and that could affect policy and election outcomes (J. Tucker et al., 2018). These distortions may be created, encouraged, and disseminated by political actors who seek to promote their policies, win an election, or avoid accountability for their actions (Fritz, Keefer, & Nyhan, 2004; Flynn, Nyhan, & Reifler, 2017).

**Filling data voids with misinformation**

"Data voids" are created when relevant data is limited, non-existent, or deeply problematic (Golebiewski, 2018). Data voids can have obvious adverse consequences in other settings relevant to the public's welfare, i.e. elections (Galston, 2017). During the 2016 presidential campaign many institutions such as mainstream media and political party organizations eroded their credibility creating a void that was filled with misinformation by an unmediated, populist nationalism and extremist voices (Galston, 2017). It has been documented how malicious actors and digital marketers run junk news factories to disseminate misinformation (Woolley & Howard, 2018). This is problematic as, according to previous work false political news are more viral than any other type of false information (Vosoughi, Roy, & Aral, 2018). If data voids are filled with misinformation created by malicious actors, this information may continue to influence people's reasoning, even if it later turns out to be incorrect (Ecker, Lewandowsky, Swire, & Chang, 2011).

## 3. Social Media Discussions around Latinos in the 2018 Midterm Election

Our data analysis is interested in responding to the question: What topics related to the Latino community were discussed before the 2018 midterm elections?

To answer this, we performed a data analysis on one of the most prominent and popular social media platforms: Reddit. We chose Reddit as they were among the platforms that played a very important role in the dissemination of information during the US presidential elections of 2016 (Kreiss & McGregor, 2018; Roozenbeek & Palau, 2017).

We conducted a content analysis over Reddit posts related to the midterm elections and Latinos. We made use of both qualitative and quantitative methods of analysis.

### 3.1. Methodology

Our goal was to understand what Latino topics were discussed or which types of accounts were mentioning Latinos during the midterm elections on Reddit.

We collected Reddit posts related to Latinos and elections from September 24, 2017 to September 24, 2018. Our data collection on Reddit consists of 1,463 unique posts and 968 unique users. The data was collected using the Reddit streaming API, which collects posts across different subreddits (communities on Reddit). We identified that the posts we collected came from six different political subreddits: r/Ask_Politics, r/Politics, r/News, r/True_News, r/Political_Humor, r/The_Donald, r/Democrats, r/Republicans. It is interesting to note that subreddits dedicated to Latino topics, such as *"r/Latino"* or *"r/mexico"* did not seem to have posts related to the midterm elections. Only the six subreddits we mentioned above had posts that directly mentioned the midterm elections and Latinos.

To identify the Reddit content related to Latinos and the midterm elections, we follow a similar approach to analyze online audiences (Stewart, Arif, Nied, Spiro, & Starbird, 2017; Savage & Monroy-Hernández, 2015). We manually created a list of keywords related to the Latino community and the midterm elections. We included synonym words that were used for "Latino" to collect posts from different social groups referencing them (e.g., some people call Latinos "beeners" to reference them in a derogatory form; Latinos from California sometimes call themselves "Chicanos". To identify keywords related to the midterm elections we collected news reports and the midterm election Wikipedia pages to identify all proper names (i.e., list of candidates, and the related people involved) and major group organizations participating in the election. We aimed for these news sites to involve all political views and inclinations to avoid bias on the information we collected. The terms were both in English and Spanish. Then, we used a combination of words to limit our search results, i.e, we queried Reddit with the term: "chicanos" and midterms. Our sampling method enables us to explore narratives involving Latino community during the period studied. However it is not necessarily representative of the broader indirect discourses about the Latino community within a political context. Specifically, our sample has likely biased our analysis towards discussion about Latinos and the midterm elections. We take care to report our findings within the limitations of this sample. Our analysis use descriptive statistics, such as post volume, this helps us to visualize how information was assembled during the period studied. We then use the products from this analysis as a component of a qualitative inquiry by drilling down and sampling individually distinctive posts to discover discursive patterns and themes.

**3.1.1. Understanding How Discussions Happen Overtime**

To understand temporal participation, we conducted quantitative analysis of posts over time. The graph in Figure 1 represent the temporal distribution of the resulting dataset along with captions for the topics in the conversations happening over time. To achieve this, we plotted the total number of posts per day. We used one year data from September 24,2017 until September 24,2017.

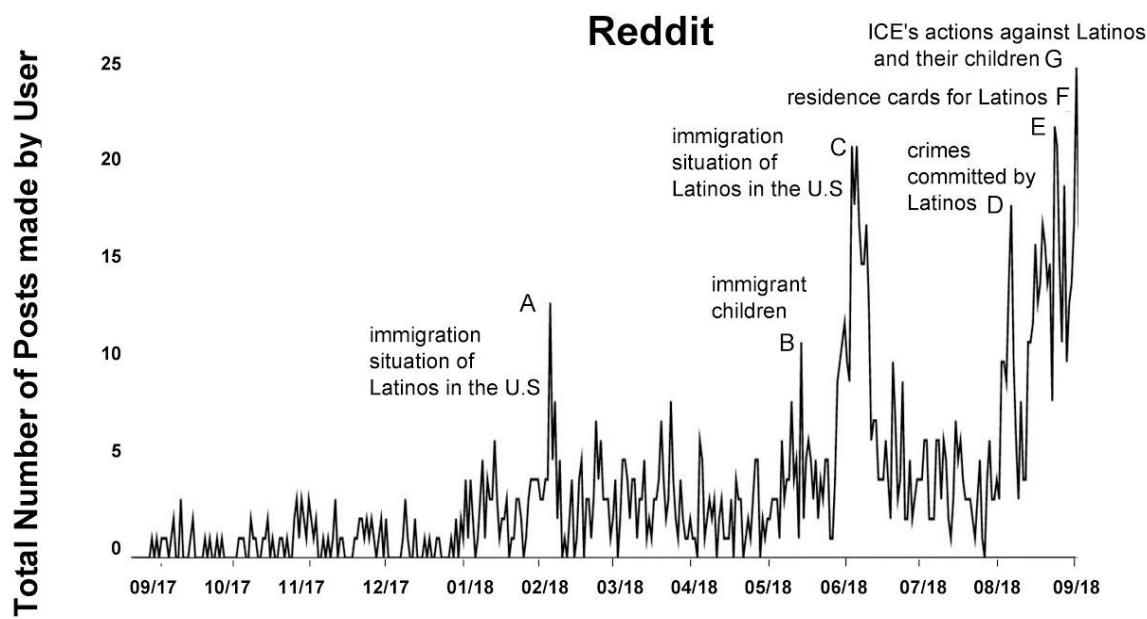

Figure 1: Overview of people's posting behavior for content related to Latinos and the 2018 midterm election on Reddit.

#### 3.1.2. Understanding Who Are the People Behind the Discussions

We use this analysis to start to uncover within our dataset, who are the main people driving conversations around Latinos and the election. For this purpose, we aimed at identifying the individuals who posted greatly about Latinos and politics, and who received wide attention from other citizens. We plotted the total number of subreddit posts per user vs its score (upvotes-downvotes) per post, see Figure 2.

### 3.2. Results

We collected one year of posts from Reddit to uncover how much people were discussing topics about Latinos on social media. We also made an effort to uncover who was driving these conversations. The data collection took a snapshot of all activity starting from September 24, 2017 until September 24, 2018.

#### 3.2.1. Understanding How Discussions Happen Overtime

The temporal distribution of the resulting dataset allows us to detect spikes of activity in the conversations we are studying on Reddit as well as the conversation relating to Latinos and midterm elections that correlated with those spikes. Figure 1 illustrates the total number of Reddit posts made per day. The X axis represents the date the post was made, and the Y axis represents the total number of posts

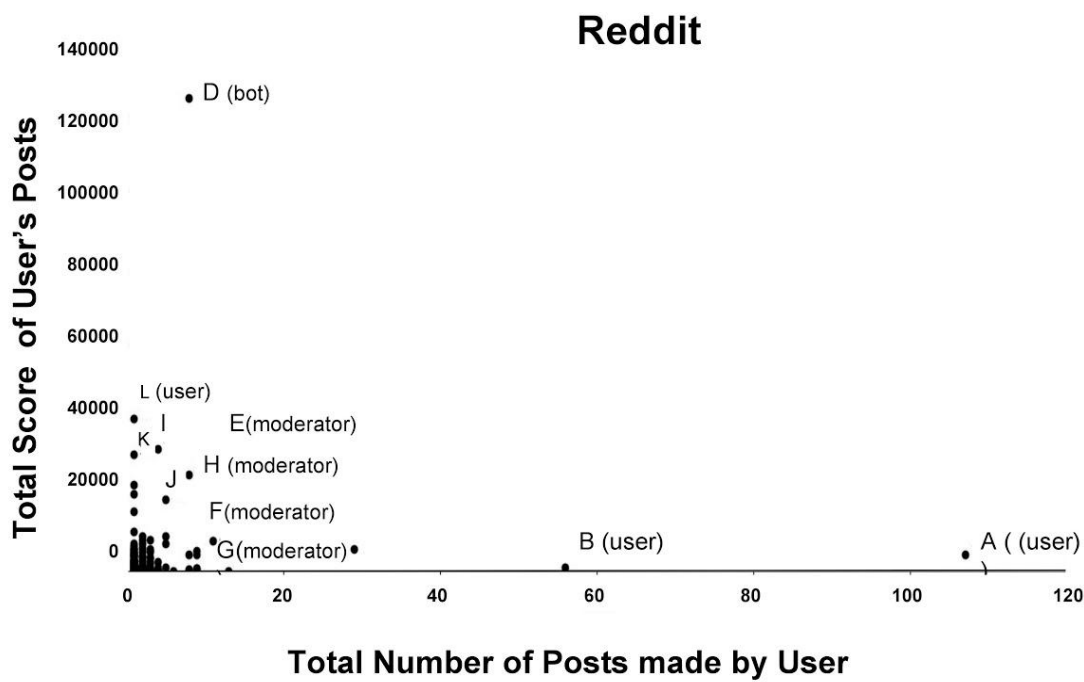

Figure 2: Overview of how much each individual person posted on Reddit and the attention they received from others.

shared that particular day.

On Reddit, we observe (see Fig.1 that posts about Latinos and elections increased very early on in the year (February 2018). The posts in February 2018 were about the primaries (Politico, 2018) that were happening that month in Texas, a state with a high number of Latinos (Dep.Com.U.S., 2010).

We noted that most of the peaks on Reddit were discussions around immigration issues. For instance on February 14, 2018 (point A), there is a spike on posts about immigration. This coincides with the open-ended debate on immigration that started on the Senate the evening before and stalled when Democrats objected to the Republican's first amendment, which would punish so-called sanctuary cities.

We also saw evidence of an immigration debate happening days before and after the Senate debate. This debate took place via mega-threads and AMA's. A mega-thread is a thread initiated by a moderator of a subreddit to organize the community and initiate discussions. AMAs (Ask Me Anything sessions) are special threads where users can ask any questions they want to celebrities and high-profile individuals. In this case, users from Reddit invited mainly politicians to participate in AMA's sessions. On May 29, 2018 (point B) the conversation revolved around the treatment of children separated from undocumented immigrant parents at the border. A topic that continued to generate debate on the platform. The spike on June 19, 2018 (point C) talked about state and federal lawmakers who were denied entry to a shelter in Florida, amid uproar over the federal government's break-up of families caught illegally crossing the U.S (Smiley, 2018). On September 11, 2018 (point D) there was a peak related to conversations around the murder of a teen in New Jersey allegedly by an illegal immigrant. Here there were also many posts discussing and encouraging others to support building the US Border wall. The top posts on September 20 and September 23,2018 (point E to point G) belonged to discussions about the policies Trump had proposed around immigrants, such as curtailing green cards for immigrants on public aid, and moving $260M from cancer research HIV/AIDS and other programs to cover the costs of taking custody of the children of immigrants.

We also note that participants had diverse dynamics to engage people in discussions, such as the mega-threads or the AMA dynamics. The discussions around political events and Latinos appeared to have been much more steady and continuous. We also note that there were not many discussions around the problems that Latinos went through to register to vote.

### 3.2.2. Understanding Who Are the People Behind the Discussions

We were also interested in uncovering who were the main people driving conversations around Latinos and the election. For this purpose, we plotted the total number of Reddit posts a particular user generated vs their popularity. The X axis in Figure 2 represents the total number of Reddit posts made by each user. The Y axis represents the total popularity of the user's posts combined (number of upvotes or favorites). Each point in Fig. 2 represents a user on Reddit.

We identified the most active users by finding those individuals whose number of posts was higher than three times the standard deviation (normal procedure to find outliers). We then profiled these active users (they are each labeled with a letter). We analyzed the type of topics that they mentioned in their content. We identified and profiled the most active users on Reddit. Here we identified three distinct behaviors in the most active:

- **User Type A ("Latino Aware + anti-Trump"):** All users in this group (16% of the highly active users) referenced the overall migratory situation of Latinos in the US primarily with an emphasis of considering president Trump a racist and anti-immigrant person.

    This behavior lead us to call this group *"Pro-Latinos + anti-Trump"*. On average, 26% of all the Reddit posts that these users made discussed how Trump was racist, and how he was cruel and inhuman by separating immigrant children from their parents at the border.

    A sample post that people from this group shared:

    > *"Gratuitous cruelty by Homeland Security: Separating a 7-year-old from her mother – WHAT EXACTLY did a 7-year-old Congolese girl do to the US to deserve the trauma that*

*has been visited upon her including forcible separation from her mother by DHS Secretary Nielsen and her immigration agents?"*

- **User Type B ("Pro-Trump + trolls")**: All authors in this group (41.5% of most active users) belonged to r/The_Donald, a community known for its political trolling behaviour (Flores-Saviaga, 2018). Their posts focused on mobilizing people to vote Republican (pro-Trump) in the midterm elections. This lead us to call this group *"Pro-Trump + trolls"*. What is interesting is that 34% of the posts generated by this group were mega-threads where they had deep discussions with others on Reddit. These mega-threads occurred at least once per week (See Figure 3). They also organized AMAs with candidates and politicians participating in the midterm elections. For instance, one of their posts read:

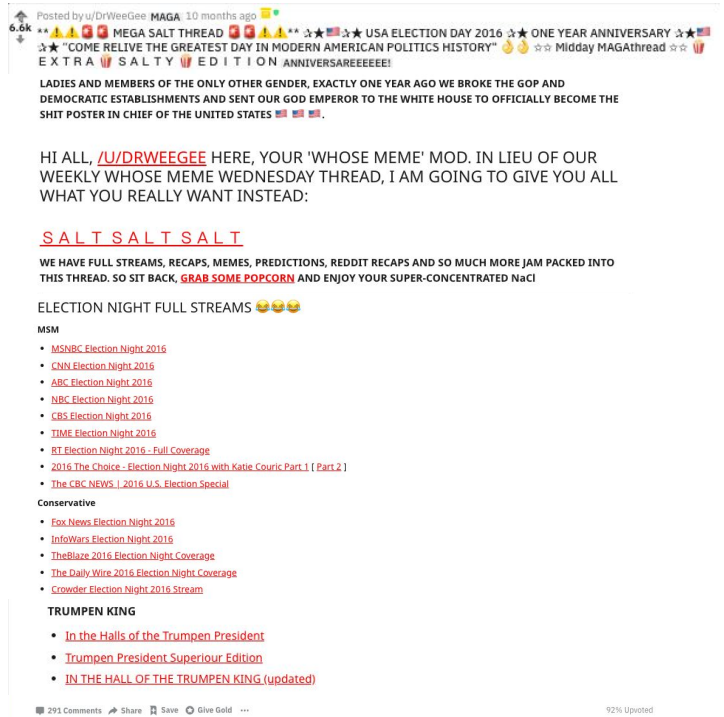

Figure 3: Example of political mega-thread from r/The_Donald where they discuss the political ecosystem, including political topics related to Latinos.

*"If you are a congressional candidate and are interested in holding an AMA(Ask Me Anything) on r/The_Congress, please contact The_Donald moderators by clicking the contact link on our sidebar"*

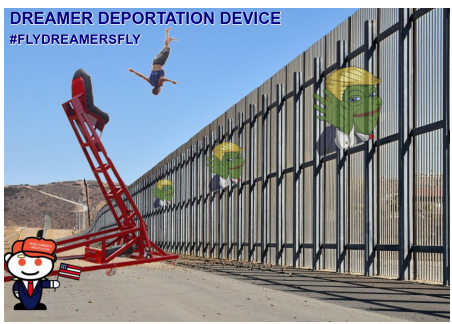

Figure 4: Meme from r/The_Donald mocking Latinos.

All of these active users posted about the current immigration situation of the United States and occasionally posted news about crimes that illegal immigrants had done. There was also a tendency to use such news reports to show special favoritism towards Trump and his decisions relating to illegal immigrant:

*"Dad's grief leads to a quest to count deaths caused by illegal immigrant drivers. In the*

> *wake of a 2013 study by the California Department of Motor Vehicles that concluded that unlicensed drivers are almost three times as likely to cause a deadly car accident as a licensed driver."*

The people in this group also occasional posted jokes about illegal immigrants (see Fig. 4).

These users also tended to post pictures of Latinos supporting Trump and encouraged Latinos to vote Republican (see Fig. 5). We also identified that within this political troll group there was a tendency to organize collective action on other platforms such as Twitter (Robertson, Vatrapu, & Medina, 2010).

- **User Type C ("The Neutrals")** This group (41.5% of the most active users) had a more neutral view on the topic of Latinos and their rights in the US They primarily posted news reports from sites that are known to have a neutral tone. For this reason, we called the groups *"The Neutrals"*. 35% of the posts of these users were mega-threads where they discussed the political ecosystem in general (without supporting Republicans or Republicans). We also noted that people in this group shared posts where fake news regarding Latinos were debunked. However, this was less than 1% of their posts. An example:

    > *"Breitbart fabricated a fake story that illegal immigrant started deadly Sonoma wildfires says Sherriff"*

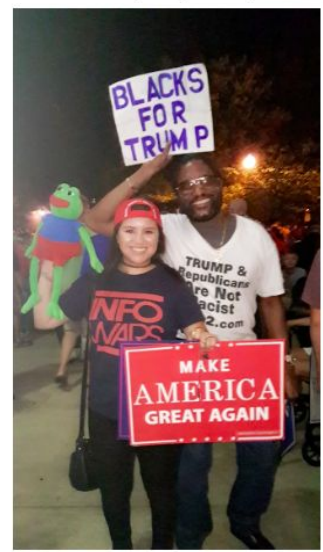

Figure 5: Example of a post on r/The_Donald.

## 4. Discussion and Future Work

Current research has focused on how political disinformation targets populations at large. However, we lacked knowledge about how the social media discussions about minorities, especially Latinos were evolving.Our research starts to address this gap by investigating how Latinos are mentioned and discussed on Reddit during the 2018 midterm elections. Through our analysis, we saw spikes in the number of posts early on in the year when people started mentioning and covering the situation of Latino families who had just arrived at the US and were being separated at the border.

We also discovered that the groups that target and covered more generally all Latino-related topics were primarily political trolls (i.e., users on the subreddit r/The_Donald, which prior work has identified as a political troll community (Flores-Saviaga, 2018)). Political trolls also had some dynamics to turn conversations around Latinos into something interesting and engaging for everyone. For instance, we found evidence of how political trolls introduced mega-threads to drive a large number of people to discuss topics around Latinos and the midterm elections. We found evidence that these mega-threads were happening at least once per week on Reddit via AMAs and mega-threads. Through these mega-threads people on Reddit appeared to contextualize, explain, and discuss in detail the political ecosystem, especially related to all Latino groups. This type of social interaction where people take the time to explain in detail to others what is happening politically is something that our research had identified is effective for engaging and mobilizing individuals to action (Flores-Saviaga, 2018). Our analysis also uncovered that political trolls created several AMA's (Ask Me Anything sessions) on Reddit with congressional candidates and political personalities where they discussed political topics regarding Latinos and the midterm elections. This type of dynamics might also help to turn the topic of Latinos and politics into something interesting and through which they can more effectively mobilize people to the polls. Future work could investigate how these interactions on social media affect people's voting behavior.

Our research also uncovered how political trolls appeared to covering and appropriating data voids around Latinos in US elections. Data voids occur when there is limited or non-existent information about certain topics (Golebiewski, 2018). For instance, there might be vast information in English about Ted Cruz, but there is less information about the topic in Spanish. As a consequence, if a person who only speaks Spanish searches for Ted Cruz to make a decision for the elections, she will likely obtain results that are not what she wants because the information is lacking.

When there is a lack of high-quality content to cover data voids, new malicious content can easily surface. For instance, if there is no content available in Spanish about Ted Cruz, a malicious actor could fill that void with the information they desire (e.g., creating fake articles in Spanish claiming that Ted Cruz has paid the college tuition of US Latinos). It is difficult for a person to realize they have come across a harmful data void, as there is not other information to refute the claim. Data voids can be exploited by those with ideological, economic, or political agendas (Golebiewski, 2018). With our research, we are observing that political trolls overall appear to have much more sophisticated techniques for creating engaging content around Latinos and US elections; while neutral actors appeared to have less strategies for engaging their audiences, creating an information void which diminishes the engagement with Latinos.

We believe our data analysis is highlighting the existence of possible data voids surrounding Latinos and US politics, and uncovering how political trolls might be occupying the space. We believe it is important to think strategically about how to address this problem to limit the number of people who encounter harmful data voids. In the following, we provide some recommendations given our findings.

### 4.1. Recommendations

Mainstream media, politicians and political organizations and have a great window of opportunity for mobilizing Latinos to be more involved in politics. Previous research has shown that taking the time to explain the political ecosystem to individuals is important to mobilize them in political causes (Flores-Saviaga, 2018). We believe it can be important for them to adopt more strategies for explaining the current political ecosystem to Latinos. Especially, to make an effort to cover all political events related to Latinos and perhaps even to adopt strategies that will facilitate deep discussions (similar to those used by political trolls). Given that Latinos are the second largest racial or ethnic group behind whites in the US (Flores, 2017), if mobilized they could potentially impact the US and create truly fairer situations for all Latinos. It might also be important to fill the data void that currently exists, as it appears that only extreme groups, such as the political trolls, are the ones covering all the political events regarding Latinos. But there could be biases in what they report, and we currently appear to lack anyone to counterattack them.

We also believe it is important that this institutions take actions to actively debunk misinformation regarding Latinos, such as fake news reports about crimes done by illegal immigrants. Given what we saw on Reddit where citizens actively debunked fake news, perhaps they could collaborate with regular citizens to have an army of people to help them identify and debunk misinformation. Here we can consider implementing crowdsourcing techniques to create effective collaborations to fight misinformation.

Our results highlighted that the r/Latino subreddit and r/Mexico subreddit were not participating in discussions around US political election (although they do have many users who are in the US.) Lack of socialization among different Latino communities from Mexico, Puerto Rico, Cuba comes as a big hindrance when trying to address the Latino community as a whole in the US. A networking system, that tries to bridge the gap between Latinos of different regions and also with different seniority in the US could help to bring their voices together. Future work could explore the design of such systems to facilitate better communication and engagement between Latino communities around politics.

## 5. Limitations

The insights this investigation provides are limited by the methodology and population we studied. Our investigation also focused on breath, rather than on depth. As a result, we do not know much about the identities of the people participating on Reddit or the organizations and their concern regarding disinformation targeted toward the Latino community in the US. Future research could involve conducting detailed interviews with actors of different organizations who have some relationship to the use of social media and/or a history of engagement in online politics and digital social life.

## 5.1. Methodological challenges

This study confronted methodological challenges that must be understood to interpret our findings correctly. The seed Reddit data we used to generate our graphs is biased (because of the Latino and midterms terms we tracked). As a result, our findings are not intended to be representative of the overall Latino community conversation. Rather, we have a portion of a particular online discourse that invokes the Latino community and midterm elections directly (i.e. mentioning the word "Latino", "midterm"). Similarly, due to the incomplete nature of our data, we cannot and do not seek to quantitatively assess the overall conversations happening on both social networks regarding the Latino community and midterm elections. Our goal is to understand a snapshot of how discussion was occurring, the topics being discussed and the main actors participating in it.

## 6. Acknowledgements